\documentclass[useAMS,usenatbib]{mn2e}

\usepackage{graphicx}                   %for PS/EPS graphics inclusion, new
\usepackage{float}
\input{epsf.sty}                        %for PS/EPS graphics inclusion, old
\input{psfig.sty}
\begin{document}

\title[Slow glitches in quark star model]
{Pulsar slow glitches in a solid quark star model}

\author[Peng, Xu]{C. Peng and R. X. Xu\\
School of Physics and Yuanpei Program, Peking University, Beijing
100871, China}

\maketitle

\begin{abstract}
A series of five unusual slow glitches of the radio pulsar B1822-09
(PSR J1825-0935) were observed over the 1995-2005 interval. This
phenomenon is understood in a solid quark star model, where the
reasonable parameters for slow glitches are presented in the paper.
It is proposed that, because of increasing shear stress as a
pulsar spins down, a slow glitch may occur, beginning with a
collapse of a superficial layer of the quark star. This layer of
material turns equivalently to viscous fluid at first, the
viscosity of which helps deplete the energy released from both the
accumulated elastic energy and the gravitation potential. This
performs then a process of slow glitch.
Numerical calculations show that the observed slow glitches could
be reproduced if the effective coefficient of viscosity is $\sim
10^2$ cm$^{2}\cdot $s$^{-1}$ and the initial velocity of the
superficial layer is order of $10^{-10}$ cm$\cdot$s$^{-1}$ in the
coordinate rotating frame of the star.
\end{abstract}

\begin{keywords}
dense matter - stars: neutron - pulsars: general - pulsars:
individual: PSR B1822-09
\end{keywords}

\section{Introduction}

It is a pity that we are now {\em not} able to determine
confidently which state of matter really exists in pulsar-like
stars due to the difficulty of calculation from the first
principles of the elementary strong interaction, even 40 years
after the discovery of pulsars.
Nuclear matter (related to {\em neutron stars}) is one of the
speculations even from the Landau's time, while quark matter
(related to {\em quark stars}) is an alternative due to the fact
of the asymptotic freedom of strong interaction between quarks
\citep[see reviews, e.g.,][]{madsenreview,glen,lp04,weber05}.
We then have to focus on astrophysical observations in order to
solve this important and fundamental question.

Actually, pulsars are ideal astro-laboratories for physics of cold
matter at supranuclear density.
Based on the Planck-like spectrum without atomic features and the
precession properties of pulsar-like stars, a solid quark matter
conjecture was suggested \citep{xu03}. Additionally, other
features naturally explained within this model could possibly
include sub-pulse drifting in radio emission \citep{xqz99},
glitches \citep{z04}, strong magnetic fields, birth after a
successful core-collapse supernova, and detection of small
bolometric radii \citep[for a short review, see, e.g.,][]{xu06}.
Besides, the observational features of anomalous X-ray
pulsars/soft gamma-ray repeaters may also reflect the nature of
solid quark stars \citep{owen05,Horvath05,xty06}.

The slow glitches recently observed in PSR B1822-09
\citep{Sh98,Sh00,zwz04,Sh05,Sh07} could be a new probe into cold
solid quark matter.
The pulsar has a period of 0.769 s and a relative young age of 230
kyr. What is interesting in this pulsar is the change in its
period known as slow glitches. Different from the typical feature
of glitch phenomenon of other pulsars, this pulsar shows glitches
that have rather slow increase in spin frequency, with a time
scale of 200-300 days (while the normal glitches experience this
process much swifter, maybe less than a spin period), and
moreover, this pulsar didn't experience any relaxation progress
during the post slow glitch days.

A series of five slow glitches are shown in Fig.1.
Fig. 1(a) shows the frequency derivative $\dot\nu$. The peaks of
$\dot\nu$ are enveloped in a parabolic curve, that may indicate
that all the slow glitches could be the components of one process.
This process could be triggered by the small glitch
($\Delta\nu/\nu\sim 8\times 10^{-10}$) occurred in 1994 September.
Both Fig. 1 (b) and (c) show the frequency residual $\Delta\nu$,
but relative to fits to data for different intervals (1991-1994
for the former, and 1995-2004 for the latter).
In this observational result, it is evident that the typical
character of slow glitch in Fig. 1(b): the spin frequency
experiences a process that increases gradually but never decreases
(i.e. no post-glitch relaxation there).
\begin{figure}
\includegraphics[width=3in]{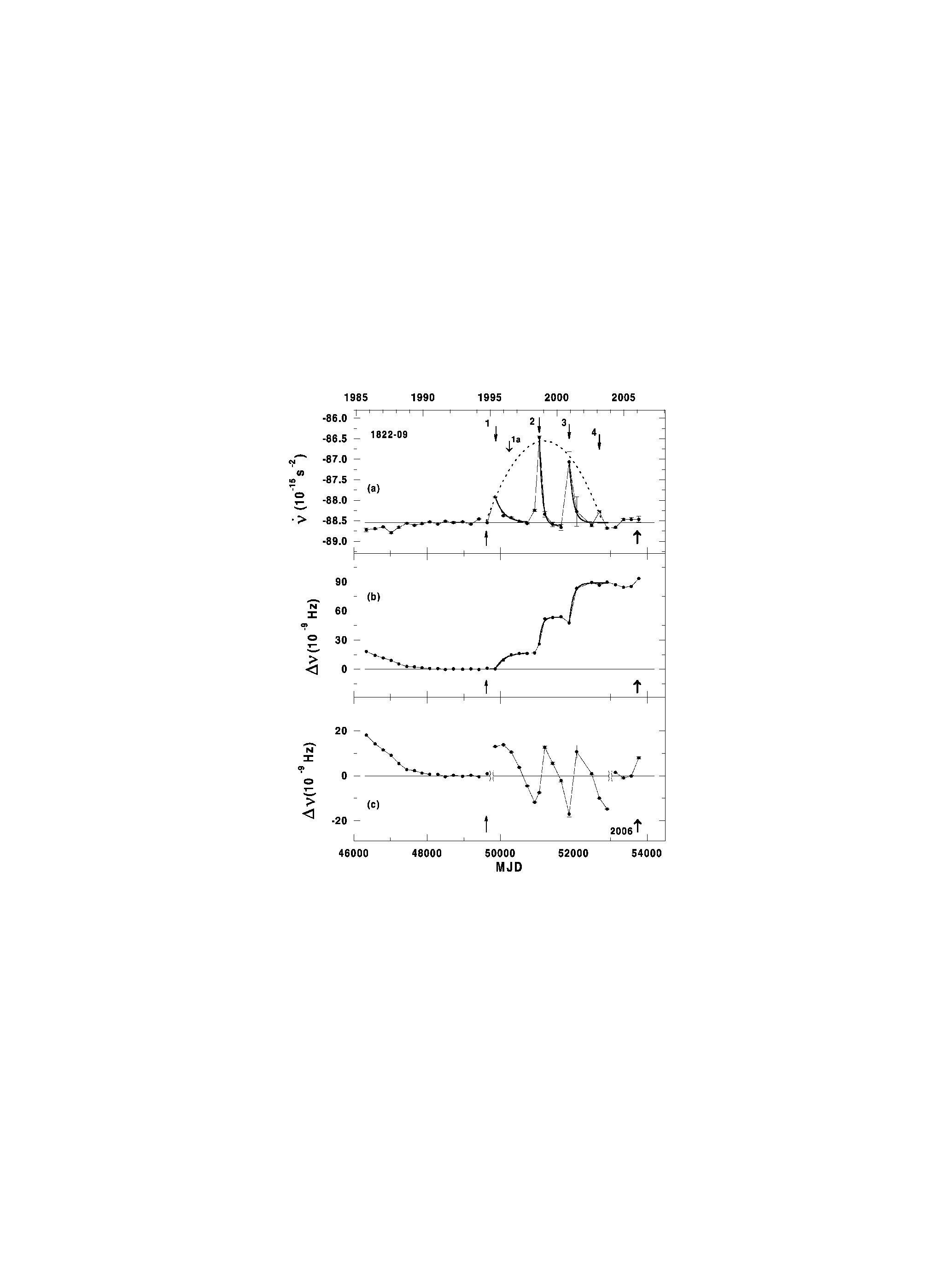}
\caption{%
The observational data of frequency residual $\Delta\nu$ and
frequency derivative $\dot \nu$ for the pulsar B1822-09. Arrows
pointing downwards indicate the epochs at which five slow glitches
occurred, while arrows pointing upwards indicate two normal
glitches. The figure is drawn by Shabanova (2007).
\label{R}}
\end{figure}

We are trying to explain this phenomenon in the regime of solid
quark stars. Since the pulsar is supposed to be mostly built up of
solid quark matter and the typical density there is order of
$10^{14}$ g/cm$^3$, soon after a quake, the motion of only a thin
layer of the quark matter may cause significant change in the moment
of inertia (and thus the spin frequency) of the star owing to the
extremely high density. Numerical calculation is carried out to
simulate the observational results of both the slow increase in the
rotating frequency and the lack of the relaxation of the post-glitch
progress.

\section{The model}

We suggest that the star is a solid quark star with a typical
density of $10^{14}$g/cm$^3$. Due to magnetodipole radiation and
particle ejection of the star, its rotation energy loses and
frequency of the rotation declines. While on the other hand,
provided that the pulsar cools down from a hotter fluid ellipse, the
equilibrium of the pulsar, with spin frequency $\nu$, results in an
eccentricity of $e$,
\begin{eqnarray}
\nu^2=8\pi^3G\rho[\frac{\sqrt{1-e^2}}{e^3}(3-2e^2)\arcsin{e}-\frac{3(1-e^2)}{e^2}],
\label{Mac}
\end{eqnarray}
where $G$ is the gravitational constant and $\rho$ is the averaged
density of the star \citep[e.g.,][]{ST83}.
The configuration of Eq.(\ref{Mac}) is for spinning stars made up
of homogenous fluid matter. Quark stars could be well approximated
by uniform density if their masses are not greater than $\sim
1.5M_\odot$ \citep{afo86}.
Eq.(\ref{Mac}) could be simplified for stars with small
eccentricities as long as they spin not very fast,
\begin{eqnarray}
\nu=4\pi e\sqrt{\frac{2\pi\rho G}{15}}. %
\label{Mac_simple}
\end{eqnarray}

The relation of Eq.(\ref{Mac_simple}) implies that when the
frequency of a star decreases, the eccentricity of the star should
decrease accordingly in order to maintain the equilibrium shape if
the star is in a fluid state.
But, as the pulsar cools down and stays as solid star, the elastic
stress might occur and accumulate to resist the changing of the
star shape. When this stress develops to a critical value that the
material of the star cannot afford, part of the star breaks (or
{\em collapses}) and then the elastic energy releases. This is
known as a quake of solid quark star \citep{z04,xu06}, during
which both elastic and gravitational energies are released.
In the model of \cite{z04}, the moment of inertia, $I$, decreases
suddenly (to result in a sharp jump of $\Delta\nu$), and then
increases gradually (to result in a following post-glitch
relaxation).
However, as elastic force develops to a critical value, a solid
rotator with smaller shear modulus might not be so violent (i.e.,
$I$ does not decrease suddenly) but in a gentle way for $I$ to
decrease, and could thus reproduce the observational feature of
slow glitches. No $I$-incensement occurs if no significant elastic
energy is converted to kinematic energy (and thus no post-glitch).
The small glitch at 1994 of PSR B1822-09 may lead to a small
effective shear modulus (e.g., significant part of the quark
matter would be shiver-like), and thus trigger the slow glitches.
A pre-glitch could then be expected for slow glitches in the
star-quake model of quark stars, in order to have an effective
small shear modulus.

Soon after a quake, we assume that a superficial layer moves in
order to set a new equilibrium (see Fig. 2), while most of the
inner part of star keeps almost the same.
Part of the energy released turns into heat and melts the debris,
while the other part turns to the kinetic energy of the fluid. It
is suggested in this model that the broken material will turn into
viscous fluid and is therefore able to flow slowly and change the
shape of the star towards the equilibrium shape.
This progress would surely cause slow decreasing of the moment of
inertia of the star. Since this process happens in a pretty short
time comparing to the typical age of the pulsar, it is reasonable
to suppose that the angular momentum of the star keeps invariant
during the whole progress of glitch.
As a result, the frequency of the star might increase slowly during
the same time. When the fluid is flowing, the viscous interaction
among the elements of the fluid may exhaust the kinetic energy of
the fluid. The shape of the star stop changing finally, and the
material then cools down and is solidified. This is why slow glitch
appears no relaxation feature in the post-glitch progress.
Calculational results (see \S3) explain why this pulsar experiences
a serious of five slow glitches rather than a single one.
\begin{figure}
\includegraphics[width=3in]{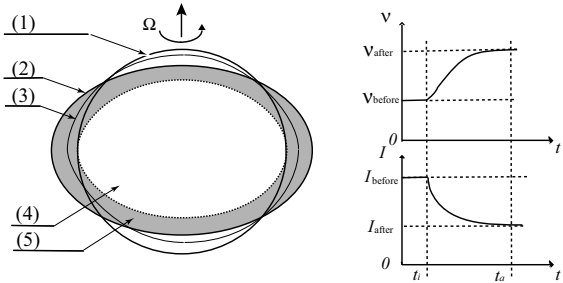}
\caption{%
Sketch of the model for slow glitches. The left side of this figure
shows the change of the eccentricity of the star during a slow
glitch and the layer activity in this progress. The right side of
this figure shows the change of $\nu$ and $I$ versus time,
respectively (Note: the usual change of $\nu$ and $I$ due to the
energy lost by magnetodipole radiation and particle emission is
neglected here).
A slow glitch is supposed to occur at time $t_i$. In the left,
(1) an imagined ellipsoidal figure (determined by Eq.(\ref{Mac}))
at $t_i$ if the star is in a fluid state, with spin frequency
$\nu=\nu_{\rm before}$;
(2) the real figure of a solid star at $t_i$ ($\nu=\nu_{\rm
before}$), with stress energy high enough to quake;
(3) the figure after slow glitch at time $t_{a}$, with
$\nu=\nu_{\rm after}$, which might be close to the imagined shape
(1);
(4) the inner solid part which keeps almost the same during a slow
glitch;
(5) the superficial crust layer which is transmuted during the slow
glitch progress.
\label{R}}
\end{figure}

To assure that model, we are doing numerical calculation to
simulate this process.
First, the Navier-Stokes equation is known as the basic equations
to describe the motion of viscous fluid. In the frame which is
fixed on the inner solid part of the star, the general form reads,
\begin{eqnarray}
\frac{\partial}{\partial
t}\textbf{v}+(\textbf{v}\cdot\nabla)\textbf{v}
=-\frac{1}{\rho}\nabla
p+\mu\Delta\textbf{v}+\textbf{g}-(\frac{d\textbf{$\omega$}}{dt}\times\textbf{r}+  \nonumber\\
+\textbf{$\omega$}\times(\textbf{$\omega$}\times\textbf{r})-2\textbf{v}\times\textbf{$\omega$}),
\label{NavSto}
\end{eqnarray}
where $\textbf{v}$ is the velocity of the superficial matter in the
rotating frame, $p$ is the pressure in the star, $\textbf{g}$ is the
acceleration of gravity on the surface of the star, $\mu$ denotes
the viscosity of the fluid after the phase transformation (from
solid to fluid-like matter), and the star is assumed to rotate
constantly. The three terms in the bracket of right hand side arises
from the inertial acceleration due to the non-inertial frame we
choose.

Let's consider briefly Eq.(\ref{NavSto}). For a zero order
approximation, a solid star could be modelled by a rigidity body,
with velocity $\textbf{v}=0$, i.e.,
\begin{eqnarray}
0 =-\frac{1}{\rho}\nabla
p+\textbf{g}-(\frac{d\textbf{$\omega$}}{dt}\times\textbf{r}+
\textbf{$\omega$} \times(\textbf{$\omega$}\times\textbf{r})).%
\label{approx}
\end{eqnarray}
The above equation could at least be adaptable for the case of
$v\ll r\omega$. Observationally, the timescale of slow glitches is
much longer than the spin period. We thus think that
Eq.(\ref{approx}) could be applied in our following simulation.
When a glitch happens, the elastic energy is released suddenly,
the solid superficial layer turns to a fluid-like state with
strong viscosity, but the pressure gradient $\nabla p$ may keep to
be almost invariant.
It is a key point that $\nabla p$ remains approximately the same
in spite of stress release in the model, which means that this is
the case in which most of the elastic energy released is changed
into heat, rather than into the kinetic energy.
The heating may melt down solid quark matter to become viscous
fluid. Soon after the starquake, the matter could be re-solidified
finally due to cooling.
Contrastively, in  case that the star breaks globally, most of the
elastic energy may turns into the form of the kinetic energy, and
the pressure gradient would change significantly during starquake;
the star would experience then a normal glitch with sharp
frequency increase.
Therefore, during a glitch, only \textbf{$\omega$} and \textbf{v}
in Eq.(\ref{NavSto}) change, with Eq.(\ref{approx}) to be well
approximated since $\Delta\omega/\omega\sim 10^{-9}$ and
$\Delta\dot{\omega}/\dot{\omega}\sim 10^{-2}$ are very small. Then
Eq.(\ref{NavSto}) of the viscous fluid can be reduced to be
\begin{eqnarray}
\frac{\partial}{\partial
t}\textbf{v}+(\textbf{v}\cdot\nabla)\textbf{v}
=\mu\Delta\textbf{v}-(-2\textbf{v}\times\textbf{$\omega$}).%
\label{dyna}
\end{eqnarray}

We are concerning the initial and boundary conditions for our
simulations.
First we assume that only part of a solid quark star breaks under
the stellar surface (the shaded region in Fig. 2), with a
thickness of $H$. We note that this assumption is for simplifying
the calculation, but is not a key for general results in the
starquake model for slow glitches (see \S3).
We denote $h$ to measure the height of matter in the beaked part,
i.e., $h=0$ in the bottom while $h=H$ on the surface.
Two cases are considered in the simulated result of Fig. 3, $H=10\%
R$ and $20\%R$, where $R$ is the stellar radius.
We suppose that the velocity of matter increases linearly from the
bottom of collapsed layer to the top, with a boundary of zero
velocity at the bottom, and that the velocity increases gradually
from polar to equator.
The tangential velocity could then be formulated as
\begin{equation}
\left\{
\begin{array}{lll}
v_{\rm max} (\theta)&=& v_i\sin \theta,\\
v(\theta, h) &=& v_{\rm max} (\theta){h\over H},%
\end{array}
\right. %
\label{initial}
\end{equation}
where \textbf{$v_{i}$} is the maximum of the initial velocity soon
after a quake. A starquake is assumed to be a trigger for the
initial movement with maximum velocity $v_i$. The continues
condition of matter (i.e. $\nabla\cdot\mathbf{v}=0$) determines
the radial velocity, given that the viscous fluid is
incompressible.

Solving Eq.(\ref{dyna}) with boundary Eq.(\ref{initial}), one can
obtain a 3-dimensional vector field of velocity all over the star.
To achieve the changing of the shape and rotating frequency, we
adopt the incompressible hypothesis and the conservation law of
angular momentum.
The movement of incompressible matter in the layer changes gradually
the shape of star, and thus the angular frequency because of
conservation of angular momentum.
In fact the conservation law of angular momentum is not obeyed
precisely due to the spindown torque. However, in light of the
rather short duration of the glitch progress, the conservation
could be well approximated during a relatively short time.

After a series of slow glitches, we may suppose that the star
stays at an equilibrium state which is described by
Eq.(\ref{Mac_simple}).
It is worth clarifying that, as noted in Fig. 2 (note the
difference between states (1) and (3)), after each slow glitch
(except the last one), the star did not turn to the real
equilibrium state, but suffer actually a series of glitches, not
only a single one.
We can then obtain the eccentricity of the star with the angular
frequency at the end of the five slow glitches through
Eq.(\ref{Mac_simple}). After doing that, we may calculate the star's
spin frequency in any time via the conservation law of angular
momentum.
%
%Actually, even if the star performs some other glitch behaviors, our
%calculation would merely be ruined out because the abnormal
%behaviors basically rely on $\triangle I/I$ or equally $\triangle
%\omega/\omega$, considering that the change in $I$ or $\omega$ is
%quite small, that might impact on our calculation result little.

Getting all above, we can carry out numerical calculus after setting
the value of initial velocity \textbf{$v_{i}$} and viscosity $\mu$.
Our goal is to find whether there is reasonable parameter space (of
\textbf{$v_{i}$} and $\mu$) where the general features of the slow
glitches observed could be reproduced. In the calculation, the
``heun'' method is adopted in order to obtain more accurate values
from numerical calculations.

\section{The results}

We do the simulation with typical density of $\rho=10^{14}$
g/cm$^3$ and stellar radius of $R=10$ km for indications. However,
according to various simulations, the general conclusions would
not change significantly if these two parameters are in the same
orders.
Observational data of slow glitches and the simulated result in
the solid quark star model are shown in Fig. 1(b) and Fig. 3,
respectively. The parameters of $\mu$ and $v_{i}$ for the five
slow glitches are also shown in table 1.

In Fig. 3, two simulated curves are shown with different thickness
of the surface fluid layer, $H=10\%R$ and $H=20\%R$, respectively.
From Fig. 3, it is obvious that the thickness indeed affect the
result of $\Delta\nu$. Though the general behavior keeps, the
amplitude and the gradient of $\Delta\nu$ increase as percentage
of the thickness to the stellar radius increases.

\begin{figure}
\includegraphics[width=3in]{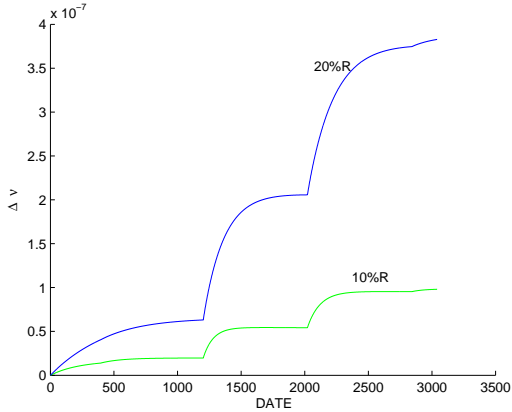}%
\caption{%
The simulation of slow glitches. The vertical axis is the
frequency change, $\Delta\nu$ (in unit of $10^{-7}$ Hz), and the
horizontal axis measures the time (in days). The two curvatures in
this figure are two simulations, given that the thickness of the
surface layer to be $10\% R$ and $20\% R$, respectively. The
starting point of time could be at MJD 49857 in Fig. 1.
\label{R}}
\end{figure}
\begin{table}
\begin{center}
  \begin{tabular}{c|c|c|c|c}
  \hline\hline\\
  { Number} & {$\mu ({\rm cm}^2\cdot {\rm s}^{-1})$} & {$v_{i}({\rm cm}\cdot {\rm s}^{-1})$}  \\
  \hline
  1& $10^{2.4}$ & $10^{-10.61}$ \\
  2& $10^{2.6}$ & $10^{-10.88}$\\
  3& $10^{2.5}$ & $10^{-10.17}$\\
  4& $10^{2.4}$ & $10^{-10.00}$ \\
  5& $10^{2.6}$ & $10^{-11.10}$ \\
\hline\hline
\end{tabular}
\end{center}
\caption{%
The parameters in the simulation where the thickness of the surface
layer is set $10\%R$. The corresponding simulated result is shown in
Fig. 3.}
\end{table}

What if the values $v_i$ of and $\mu$ in Table 1 change?
In case of $H=10\%R$, the influence of these parameters on the
simulated result of $\Delta\nu$ are shown in Fig. 4 and Fig. 5,
respectively.

As shown in Fig. 4, where values of $\mu$ are the same listed in
Table 1, $\Delta\nu$-curves for various $v_{i}$ are illustrated.
It is evident that a larger value of $v_{i}$ could result in a
larger glitch amplitude of $\Delta\nu$. It is worth noting,
however, that the time durations in which $\Delta\nu$ increases to
the maximum are almost the same. This could be easily understood
since a larger $v_{i}$ implies a larger stress and energy released
in the starquake.
The effect of changing $\mu$-values on the amplitude of
$\Delta\nu$ are demonstrated in Fig. 5, where the values of
$v_{i}$ are the same listed in Table 1.
As shown in Fig. 5, smaller $\mu$-values could not only result in
higher amplitudes of $\Delta\nu$, but also in longer durations for
$\Delta\nu$ to increase to maximum.
This result is also understandable. In the model, the fluid moves
under coriolis' force and viscous resistance. Smaller $\mu$ could
lead to a relative smaller resistance and thus a longer time to
reach its equilibrium.
In summary, one needs to have a high $v_i$ or a low $\mu$ in order
to reproduce a large amplitude of $\Delta\nu$, while a small
$\nu$-value could effectively delay the increase of $\Delta\nu$.

\begin{figure}
\includegraphics[width=3in]{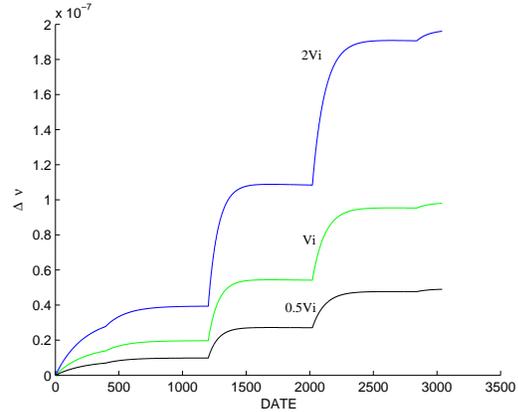}%
\caption{%
Different curves of $\Delta\nu$ with different parameters of
$v_{i}$. The curve labelled as ``Vi'' is that same one labelled as
``$10\%$R'' in Fig. 3, with $v_i$-values listed in Table 1. The
curves labelled as ``2Vi'' and ``0.5Vi'' are with $v_i$-values
double and half the ones listed in Table 1, respectively.
\label{R}}
\end{figure}

\begin{figure}
\includegraphics[width=3in]{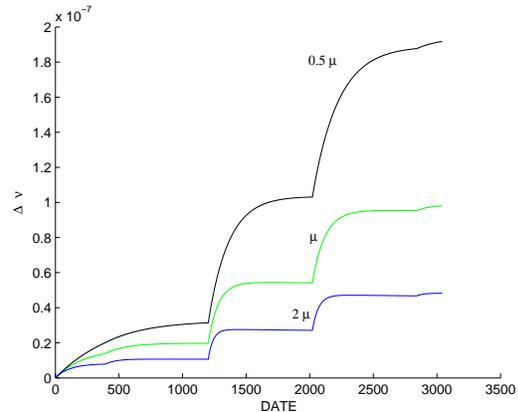}%
\caption{%
Different curves of $\Delta\nu$ with different parameters of
$\mu$. The curve labelled as ``$\mu$'' is that same one labelled
as ``$10\%$R'' in Fig. 3, with $\mu$-values listed in Table 1. The
curves labelled as ``$2\mu$'' and ``$0.5\mu$'' are with
$\mu$-values double and half the ones listed in Table 1,
respectively.
\label{R}}
\end{figure}

\section{Conclusions and Discussions}

A solid quark star model, in which the surface matter breaks during
an initial small glitch, is suggested for understanding the general
features of slow glitches observed recently.
Though this is a real problem of rheology in principle, we deal
with the matter as viscous fluid for simplicity.
Numerical calculations show that the observed slow glitches could
be reproduced if the effective coefficient of viscosity is $\sim
10^2$ cm$^{2}\cdot $s$^{-1}$ and the initial velocity of the
superficial layer is order of $10^{-10}$ cm$\cdot$s$^{-1}$ in the
coordinate rotating frame of the star.
We simulated with typical density of $\rho=10^{14}$ g/cm$^3$ and
stellar radius of $R=10$ km for indications, but the general
conclusions would not change significantly if these two parameters
are in the same orders.

The pieces of blocks should be re-correlated through solidifying
melted surfaces of the pieces, into which the superficial layer of
a solid quark star breaks soon after a quake.
The released energy (both gravitational and elastic) during a slow
glitch may result in melting the conjunctional parts of segments
(i.e., the surfaces of blocks) at first. The temperature then
increases significantly at only a small fraction ($\eta\ll 1$) of
the segments. For an order-of-magnitude demonstration, we apply the
Debye model to normal solid quark matter and obtain a heat capacity
of
\begin{equation}\label{R}
    C_{\rm
    v}=0.15\frac{k_{B}^{4}R^{3}T^{3}}{c^{3}\hbar^{3}}\frac{\rho_{0}}{\rho_{B}}\eta.
\end{equation}
The matter absorbs an energy of $E=\int_0^T{C_{\rm v}{\rm d}T}$ from
an initial temperature of $T_0\ll T$ to $T$.
In case of released energy of $E\sim 10^{39}$ erg and a melting
temperature of $T\sim 10$ MeV ($\gg$ the star's global temperature
$T_{0}\sim$ keV), we have,
\begin{equation}\label{R}
    \eta=2.7\times10^{-22}\frac{c^{3}\hbar^{3}}{k_{B}^{4}R_{6}^{3}T_{11}^{4}}\frac{\rho}{\rho_{0}}\approx10^{-8}\frac{1}{R_{6}^{3}T_{11}^{4}}\frac{\rho}{\rho_{0}},
\end{equation}
for stars with radius of $R_6\times 10^6$ cm and density of $\rho$,
where $T_{11}\times 10^{11}$ K is the melting temperature and
$\rho_0$ is the nuclear density.
For blocks with length scale of 1 m, this calculation shows a
heating surface part with thickness of about $10^7$ fm for each of
the blocks. We note the timescale for keeping such a high
temperature gradient should be very small, so that the contacting
parts cool rapidly to solidify. More and more small blocks become
correlated (i.e., a big solid-like bulk matter forms) as a slow
glitch evolves, and strain energy develops again then.
All the heated points are within the star and the cold surface
keeps, and then no significant heating feature could be observed
after a glitch.

Alternatively, let's consider a crusted strange quark star with a
solid crust and a fluid quark matter core. The inertia of the
crust could be only $\sim 10^{-5}$ times of that of the quark
matter core. According to ${\triangle I/I}\sim {\triangle
\Omega/\Omega}\sim 10^{-8}$ for slow glitches, the crust
ellipticity should change a value of $\sim 10^{-3}$, while the
actual ellipticity is $\sim 10^{-5}$ for stars with period of
$\sim 1$ s.
This means that pulsar slow glitches could not be re-produced in a
crusted strange star model.

What is the key ingredient which affects a pulsar to undergo a
normal or a slow glitch?
We think that the stellar mass of quark star could play an
important role in this issue.

Let's introduce two kinds of stress force inside solid stars at
first.
As noted by \cite{xu06}, two kinds of factors could result in the
development of stress energy in a solid star, and then in star
quakes as glitches.
(i) As a quark star cools (even spinning constantly), changing
state of matter may cause a development of anisotropic pressure
distributed inside a solid matter. Such matter cannot be well
approximated by perfect fluid, and the equation governing star's
gravitational equilibrium should then not be the TOV equation. In
case of spherical symmetry (the simplest case), one can introduce
the difference between radial and tangential pressures, $\Delta$.
Change of $\Delta$ would lead to no-conservation of stellar
volume. We call the force, which acts primary in this case, as
{\em bulk-variable force}. This kind of force may be the key
factor for giant quakes during superflares of soft $\gamma$-ray
repeaters \citep{xty06}.
(ii) An uniform fluid star would keep its eccentricity presented
in Eq.(\ref{Mac}) or Eq.(\ref{Mac_simple}), i.e., the eccentricity
decreases as a star spins down. However, for a solid star, the
shear stress would prevent the star from decreasing eccentricity
during spindown. In this case, even the state of matter does not
change, stress energy could still develop as solid star spins
down. We call this kind of force as {\em bulk-invariable force}
since the total stellar volume may always keep constantly.
Starquakes due to bulk-invariable force, and the consequent
glitches, were calculated and discussed in \citep{z04}.

Both bulk-invariable and bulk-variable forces could result in
decreases of moment of inertia, and therefore in pulsar glitches.
These two kinds of forces could trigger normal glitches if they
are relatively stronger than the critical stress, but might only
conduce to slow glitches if weaker.
The bulk-variable force could be stronger than the bulk-invariable
one for massive solid stars, since, in a special case of
non-rotating, the gravity $\propto M^2/R^2$ ($M$: mass, $R$:
radius) is strong there.
If the critical stress of solid quark matter is almost the same
(note: the effective critical stress could be much small for a
shiver-like surface layer), we may think that normal glitches
prefer to occur in massive quark stars with possibly strong
bulk-variable forces, while slow glitches are likely to take place
in low-mass solid quark stars.

What if PSR B1822-09 is a low-mass quark star?
The spindown due to magnetodipole torque for a star with a
magnetic dipole moment $\mu$, a moment of inertia $I$, and an
angular velocity $\Omega$ reads
\begin{equation}
{\dot \Omega} = -{2\over 3Ic^3}\mu^2\Omega^3.%
\label{dotO}
\end{equation}
This rule keeps quantitatively for any oblique rotators, as long
as the braking torques due to magnetodipole radiation and the
unipolar generator are combined \citep{xq01}.
Approximating $I= (2/5)MR^2$, $\mu=(1/2)BR^3$, and
$M=(4\pi/3)R^3\rho$, with $\rho$ the average density, we have then
\begin{equation}
R = 2.76\times 10^{31}\rho_{14}B^{-2}~~{\rm cm}.%
\label{R(B)}
\end{equation}
where $\rho=\rho_{14}\times 10^{14}$g/cn$^3$, and $P=0.769$ s and
${\dot P}=5.23\times 10^{-14}$ for PSR B1822-09. The pulsar could
be a few kilometers in radius if the polar magnetic field is $\sim
10^{13}$ G.

The potential drop in the open field line region, $\phi$, should
be greater than a critical value of $\sim 10^{12}$ V $\simeq
3\times 10^9$ e.s.u. in order to create secondary $e^\pm$ plasma
via gap discharges. The potential drop between the center and the
edge of a polar cap is therefore \citep{rs75}%
\begin{equation}
\phi={2\pi^2\over c^2}R^3BP^{-2}=7.84\times 10^{74}\rho_{14}^3B^{-5}.%
\label{phi}
\end{equation}
Note that the potential in above equation would be higher if the
effect of inclination angle is included \citep{ycx06}. The drop
could be high enough for pair production if the field $B\sim
10^{13}$ G, and the pulsar should then be radio loud.

%%%%%%%%%%%%%%%%%%%%%%%%%%%%%%%%%%%%%%%%%%%%%%%%%%%%%%%%%%%%%%%%%%%%%
\section*{Acknowledgments}
\thanks{%
The authors thank Y. L. Yue and others in the pulsar group of
Peking University for helpful discussions and an anonymous referee
for valuable comments. This work is supported by NSFC (10573002,
10778611), the Key Grant Project of Chinese Ministry of Education
(305001), the program of the Light in China's Western Region
(LCWR, No. LHXZ200602), and the President's Undergraduate Research
Fellowship of Peking University.}
%%%%%%%%%%%%%%%%%%%%%%%%%%%%%%%%%%%%%%%%%%%%%%%%%%%%%%%%%%%%%%%%%%%%%

\end{document}